\documentclass{emulateapj}
\usepackage{apjfonts}
\usepackage{graphicx}
\usepackage{amsmath}
\usepackage{amsfonts}

\shortauthors{Sanchis-Ojeda et al.~2010}
\shorttitle{Rotation angles of WASP-4b}

\begin{document}

%
\def\ltsima{$\; \buildrel < \over \sim \;$}
\def\lsim{\lower.5ex\hbox{\ltsima}}
\def\gtsima{$\; \buildrel > \over \sim \;$}
\def\gsim{\lower.5ex\hbox{\gtsima}}
%

\bibliographystyle{apj}

\title{
  Starspots and spin-orbit alignment
   in the WASP-4 exoplanetary system$^\star$
}

\author{
Roberto Sanchis-Ojeda\altaffilmark{1},
Joshua N.\ Winn\altaffilmark{1},
Matthew J.\ Holman\altaffilmark{2},\\
Joshua A.\ Carter\altaffilmark{1,2,3},
David J.\ Osip\altaffilmark{4},
Cesar I.\ Fuentes\altaffilmark{2,5}
}

\altaffiltext{$\star$}{Based on observations with the 6.5m Magellan
  Telescopes located at Las Campanas Observatory, Chile.}

\altaffiltext{1}{Department of Physics, and Kavli Institute for
  Astrophysics and Space Research, Massachusetts Institute of
  Technology, Cambridge, MA 02139, USA}

\altaffiltext{2}{Harvard-Smithsonian Center for Astrophysics, 60
  Garden Street, Cambridge, MA 02138, USA}

\altaffiltext{3}{Hubble Fellow}

\altaffiltext{4}{Las Campanas Observatory, Carnegie Observatories,
  Casilla 601, La Serena, Chile}

\altaffiltext{5}{Department of Physics and Astronomy,
  Northern Arizona University, P.O.\ Box 6010, Flagstaff, AZ 86011}

\begin{abstract}

  We present photometry of 4 transits of the exoplanet WASP-4b, each with a precision of approximately 500 ppm and a time sampling of 40-60~s. We have used the data to refine the estimates of the system parameters and ephemerides. During two of the transits we observed a short-lived, low-amplitude anomaly that we interpret as the occultation of a starspot by the planet. We also found evidence for a pair of similar anomalies in previously published photometry. The recurrence of these anomalies suggests that the stellar rotation axis is nearly aligned with the orbital axis, or else the star spot would not have remained on the transit chord.  By analyzing the timings of the anomalies we find the sky-projected stellar obliquity to be $\lambda = -1_{-12}^{+14}$~degrees. This result is consistent with (and more constraining than) a recent observation of the Rossiter-McLaughlin effect. It suggests that the planet migration mechanism preserved the initially low obliquity, or else that tidal evolution has realigned the system.  Future applications of this method using data from the {\it CoRoT} and {\it Kepler} missions will allow spin-orbit alignment to be probed for many other exoplanets.

\end{abstract}

  
\section{Introduction}
  
Spots on the host stars of transiting planets have generally been regarded as a nuisance. They interfere with the determination of the planet's properties, by causing variations in the transit depth, producing chromatic effects that can be mistaken for atmospheric absorption, and causing anomalies in individual light curves when spots are occulted by the planet (see, e.g., Rabus et al.~2009, Knutson et al.~2009, Carter et al.~2011).

Silva-Valio~(2008) pointed out that starspots may be helpful in one respect: observations of spot-occultation anomalies in two closely-spaced transits can be used to estimate the stellar rotation period. In effect, the planet is used to reveal the longitude of the spot during each transit. For the particular case of CoRoT-2, Silva-Valio et al.\ (2010) used this method to estimate the rotation period and study the distribution, shape and intensity of the spots. Likewise, Dittmann et al.~(2009) estimated the rotation period of TrES-1 using starspot anomalies.

In this paper we show how the recurrence (or not) of starspot anomalies can also be used to test whether the stellar rotation axis is aligned with the planet's orbital axis. Specifically, starspot anomalies are an alternative means of measuring or bounding $\lambda$, the angle between the sky projections of the angular momentum vectors corresponding to stellar rotation and orbital motion. The spot modeling of Silva-Valio et al.\ (2010) and Dittmann et al.~(2009) was restricted to values of $\lambda$ that were permitted by prior observations of the RM effect, but as we will show, it is possible to obtain tighter constraints on $\lambda$ using only spot anomalies.

As many authors have pointed out, measurements of stellar obliquities are important clues about the processes of planet formation, migration, and subsequent tidal evolution (see, e.g., Queloz et al.~2000; Ohta et al.~2005; Winn et al.~2005, 2010a; Fabrycky \& Winn 2009; Triaud et al.~2010; Morton and Johnson 2011).  The other main method for measuring $\lambda$ is the Rossiter-McLaughlin (RM) effect, an anomalous Doppler shift that is observed during transits due to the partial eclipse of the rotating star (see, e.g., Queloz et al.~2000, Ohta et al.~2005, Gaudi \& Winn 2007). Knowledge about spin-orbit alignment can also be gained from statistical studies of projected rotation rates (Schlaufman 2010), asteroseismology (Wright et al.~2011), and interferometry (Le~Bouquin et al.~2009).

The particular system studied here is WASP-4b, a giant planet discovered by Wilson et al.\ (2008) that transits a G7V star with a period of 1.34 days. Refined parameters for this system were presented by Winn et al.\ (2009), Gillon et al.\ (2009), and Southworth et al.\ (2009).  Observations of the RM effect by Triaud et al.~(2010) revealed the orbit to be prograde but gave only weak constraints on the projected obliquity: $\lambda = -4^\circ\, ^{+43^\circ}_{-34^\circ}$.

This paper is organized as follows. In Section 2 we report on observations of four transits of WASP-4b. In Section 3 we identify the anomalies that are interpreted as spot-crossing events, and use the remaining data to compute new system parameters.  In Section 4 we model the light curves by taking the star spot to be a circular disk with a lower intensity than the surrounding photosphere. In Section 5 we determine $\lambda$ using a simpler geometrical model, which does not make strong assumptions about the size or shape of the spots. Finally, in Section 6 we discuss the results and possible future applications of this method.

\begin{deluxetable*}{lcccccc}

\tablecaption{Observations of WASP-4\label{tbl:log}}
\tablewidth{0pt}

\tablehead{
\colhead{Date} &
\colhead{Epoch} &
\colhead{Number of} &
\colhead{Median time}&
\colhead{Airmass}&
\colhead{RMS residual}&
\colhead{Estimated Poisson} \\
\colhead{[UT]} &
\colhead{} &
\colhead{data points} &
\colhead{between points [s]} &
\colhead{} &
\colhead{[ppm]} &
\colhead{noise [ppm]}
}

\startdata
2009 Aug 02 & 260 & 369 & 56 & $1.48 \rightarrow 1.02 \rightarrow 1.11$ & 442 & 316 \\
2009 Aug 06 & 263 & 406 & 56 & $1.48 \rightarrow 1.02 \rightarrow 1.21$ & 452 & 315 \\
2009 Aug 10 & 266 & 365 & 55 & $1.34 \rightarrow 1.02 \rightarrow 1.30$ & 487 & 318 \\
2009 Sep 26 & 301 & 355 & 41 & $1.41 \rightarrow 1.02 \rightarrow 1.03$ & 588 & 373
\enddata

\end{deluxetable*}

\section{Observations and Data Reduction}

We observed the transits of UT 2009 August~02, 06 and 10, and also 2009~September~26, with the Magellan (Baade) 6.5m telescope at Las Campanas Observatory in Chile. We used the Raymond and Beverly Sackler Magellan Instant Camera (MagIC) and its SITe $2048 \times 2048$ pixel CCD detector, with a scale of 0$\farcs$069 pixel$^{-1}$. At the start of each night, we verified that the time stamps recorded by MagIC were in agreement with GPS-based times to within one second. To reduce the readout time of the CCD from 23~s to 10~s, we used the same technique used by Winn et al.~(2009): we read out a subarray of $2048\times 256$ pixels aligned in such a manner as to encompass WASP-4 and a nearby bright comparison star of similar color. The telescope was strongly defocused to spread the light over many pixels,
thereby allowing for longer exposures without saturation and reducing the impact of natural seeing variations. On each night we obtained repeated $z$-band exposures of WASP-4 and the comparison star for about 5~hr bracketing the predicted transit time. Autoguiding kept the image registration constant to within 10 pixels over the course of each night.

\begin{figure*}[ht]
\begin{center}
\leavevmode
\hbox{
\epsfxsize=7.0in
\epsffile{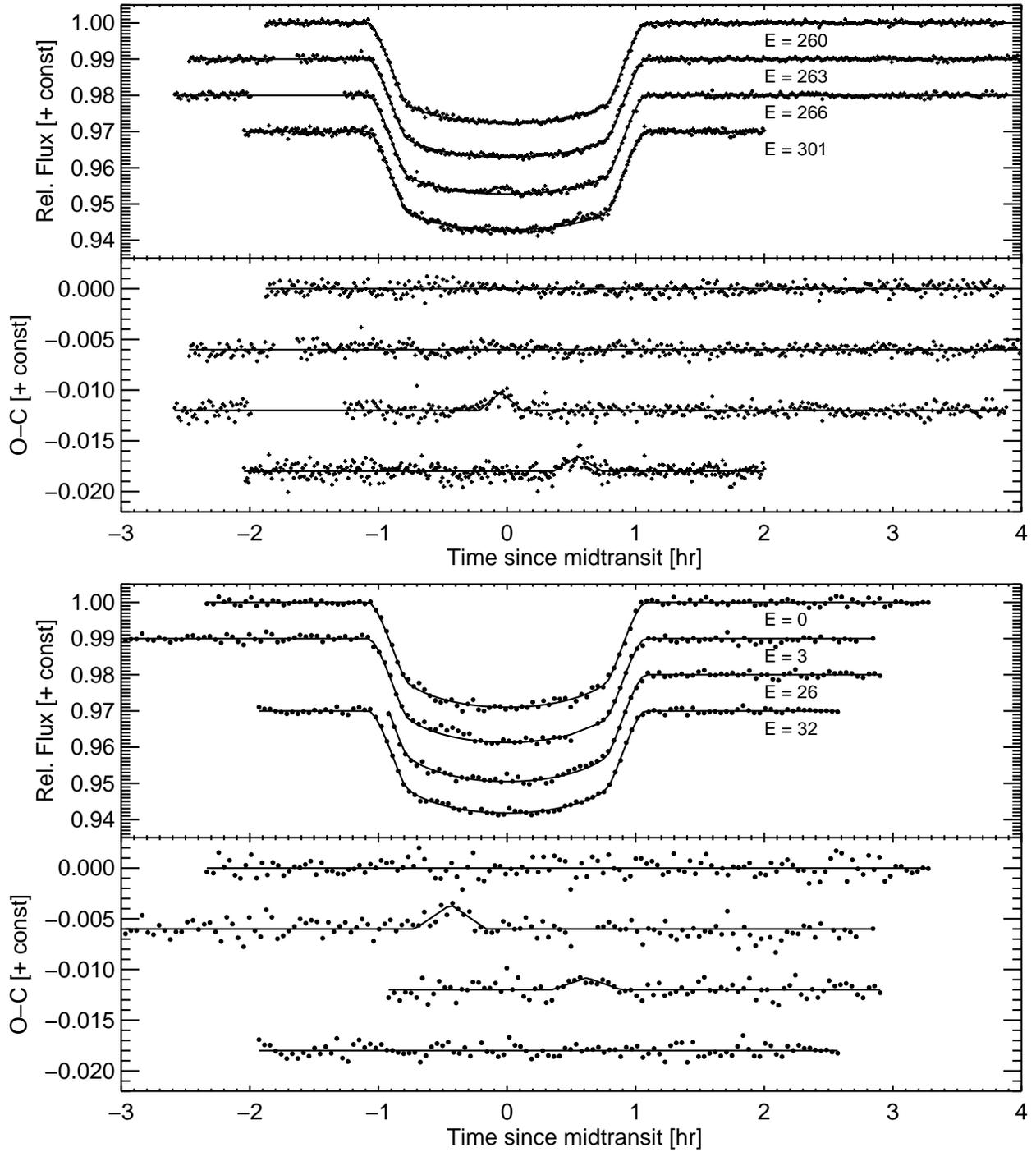}}
\end{center}
\vspace{0.00in}
\caption{WASP-4 transit light curves and starspot anomalies. {\it Upper panel}: Four different transits observed in the $z$-band with the Magellan/Baade 6.5m telescope. The solid curve shows the best-fitting transit model.  The bottom two transits display anomalies in the residuals that we interpret as spot-crossing events.  The residuals are shown below, with curves representing a simplified spot model (see Section~\ref{sec:model-geometric}).  {\it Lower panel}: A similar presentation of the four $R$-band transit light curves presented by Southworth et al.~(2009).}
\vspace{0.0in}
\label{fig:magtrans}
\end{figure*}

On the first, second, and fourth nights the skies were nearly cloud-free. The third night was partly cloudy for a short duration, and the data from that time range were disregarded. In all cases the observations bracketed the meridian crossing of WASP-4 and the maximum airmass was 1.5. We used custom IDL procedures for overscan correction, trimming, flat-field division and photometry. The flat field function for each night was calculated from the median of 80-100 $z$-band exposures of a dome-flat screen. We performed aperture photometry of WASP-4 and the comparison star, along with annular sky regions surrounding each star. Then we divided the flux of WASP-4 by the flux of the comparison star. Trends in the out-of-transit (OOT) data were observed and attributed to color-dependent differential extinction,
for which a correction was applied in the form
\begin{equation}
\Delta m_{\rm cor} = \Delta m_{\rm obs} + \Delta m_0 + k z
\end{equation}
where $z$ is the airmass, $\Delta m_{\rm obs}$ is the observed magnitude difference between the target and comparison star, $\Delta m_{\rm cor}$ is the corrected magnitude difference, $\Delta m_0$ is a constant and $k$ is the coefficient of differential extinction. Table~\ref{tbl:log} is a summary of the observations, including the standard deviation of the OOT flux, and the theoretical Poisson noise. Table~\ref{tbl:data} gives the final time series. Figure~\ref{fig:magtrans} shows the light curves, along with four light curves published previously by Southworth et al.~(2009).

\section{Starspots and System Parameters}
\label{sec:starspots_and_system_parameters}

The Magellan light curves are well-fitted by a standard transit model except for two anomalies that are visible in the third dataset ($E=266$, $t\approx -0.05$~hr from midtransit) and the fourth dataset ($E=301$, $t\approx +0.55$~hr). Each anomaly is interpreted as the temporary brightening of the system as the planet moves away from an unspotted portion of the stellar disk and onto a starspot.  Because the starspot is relatively cool and dark compared to the surrounding photosphere, the fractional loss of light due to the planet is temporarily reduced and the received flux slightly rises. The amplitude of the anomalies (about 0.1-0.2\%) corresponds to the fractional loss of light due to the starspot, i.e., the fractional area of the starspot multiplied by the intensity contrast relative to the surrounding photosphere.

The first step in our analysis was to excise the anomalous data and use the rest of the data to update the basic system parameters. For this purpose we fitted the four new data sets simultaneously with the two datasets presented by Winn et al.\ (2009), which were obtained with the same telescope and instrument. We used Mandel \& Agol's (2002) model with a quadratic limb-darkening law. We assumed the orbit to be circular, since no eccentricity has been detected with any of the existing radial-velocity data (Wilson et al.~2008, Madhusudhan \& Winn 2009, Pont et al.~2011) or occultation data (Beerer et al.~2011). There were 30 adjustable parameters: 6 midtransit times, 6 transit depths (since unocculted starspots may cause variations in transit depth), 2 limb-darkening coefficients, the impact parameter ($b$), the stellar radius in units of the orbital distance ($R_\star/a$), and 2 parameters per time series for the differential extinction corrections.\footnote{Following Winn et al.~(2009), we consider the two disjoint segments of the 2008~August~19 observation as two separate time series, for a total of 7 time series.} We refer the reader to the
description by Winn et al.~(2009) for a detailed explanation of the parameter estimation method, which is based on the Monte Carlo Markov Chain technique. The procedure takes correlated noise into account using the ``time-averaging'' method, in which the ratio $\beta$ is computed between the standard deviation of time-averaged residuals, and the standard deviation one would expect assuming white noise. This method gave values of $\beta=1.26$, 1.15, 1.00, and 1.39 for the four new light curves.

The best-fitting light curves are shown in Figure~\ref{fig:magtrans}, and the results for the parameters are in Tables~\ref{tbl:individual} and~\ref{tbl:params}. All the results for the parameters agree with the previously published values. The theoretical limb darkening coefficients obtained from Claret (2004) are $u_1=0.25$ and $u_2=0.31$, which are about 2$\sigma$ away from our results. The data prefer a smaller center-to-limb variation (smaller $u_1+u_2$) than the tabulated limb-darkening law. The six individual transit depths [i.e., the individual values of $(R_p/R_\star)^2$] had a mean of 0.02386 and a standard deviation of 0.00029, as compared to 1$\sigma$ uncertainties of about 0.00014. This suggests that the transit depth is variable at the level of $\approx$0.00025 or 1\%.  Such variations could be produced by starspots that are not necessarily on the transit chord. During each transit, a different pattern of starspots may appear on the visible hemisphere of the star, causing variations in the fractional loss of light due to the planet. Since the light-curve anomalies implicate individual spots with a fractional loss of light of only 0.1-0.2\%, the observed transit depth variations of $\approx$1\% would have to be caused by larger individual spots, or multiple spots.

The detection of the two anomalies in the Magellan data prompted us to search for similar anomalies in previously published data.  The only sufficiently precise light curves we found were the single $z$-band light curve presented by Gillon et al.~(2009), which does not display any obvious anomalies; and the four $R$-band light curves by Southworth et al.~(2009), two of which display anomalies similar to those we found in the Magellan data.  All four of the Southworth et al.~(2009) light curves are shown in Figure~\ref{fig:magtrans}.  Compared to the Magellan data, the $R$-band data have a scatter that is 40\% larger and a sampling rate three times slower, but anomalies can still be seen in the second dataset at $t=-0.4$~hr and (less obviously) in the third dataset at $t=0.6$~hr. Southworth et al.~(2009) also noted these anomalies and the possibility that they were caused by starspot occultations.

\begin{figure}[ht]
\epsscale{1.0}
\plotone{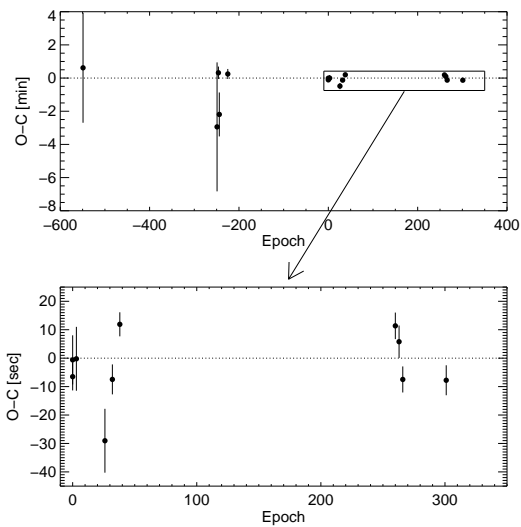}
\caption{ {\it Upper panel}: Transit timing residuals for all 15 midtransit times based on this work
and others in the literature. {\it Lower panel}: close-up of the data from the last two years,
where the excess of scatter is more noticeable due to the smaller uncertainties.}
\label{fig:timings}
\end{figure}

To refine the transit ephemeris, and search for any departures from strict periodicity, we fitted the midtransit times with a linear function of epoch. Before doing so we checked on the robustness of the uncertainties by employing an alternative technique, a bootstrap method based upon cyclic permutations of the residuals. The differences between the two methods of estimating uncertainties were no greater than 20\%. To be conservative, the ephemeris was computed using the larger of the two uncertainty estimates. The uncertainties quoted in Table~\ref{tbl:params} also represent the larger uncertainties. Figure~\ref{fig:timings} shows the observed minus calculated (O$-$C) midtransit times. The best fit to the 6 Magellan transit times gives $\chi^2 = 20$ with 4 degrees of freedom.  When we also included the other 9 data points reported by Southworth et al.~(2009),\footnote{To place all the data onto the same time standard, we used the code by Eastman et al.~(2010) to convert HJD$_{\rm UTC}$ to BJD$_{\rm TDB}$.} we found $\chi^2= 34.96$ with 13 degrees of freedom.

The probability of obtaining such a large $\chi^2$ with only random Gaussian noise is only $0.08\%$. There appears
to be a scatter of 5-10~seconds in excess of the measurement uncertainties. One possibility is that the transiting planet's orbit is being perturbed by the gravity of another planet or satellite. Another possibility is that the light curves are affected by low-level starspot anomalies (not visually recognized and excised) which are biasing the estimates of the midtransit times.

The order-of-magnitude of the apparent timing anomalies caused by occulted spots can be estimated as follows. We write the observed light curve as $1 - \delta(t) + \delta_s(t)$, where $\delta(t)$ is the fractional loss of light due to the planet, and $\delta_s(t)$ is the anomaly due to the occultation of a starspot.
Then the shift in the centroid of the light curve due to the spot anomaly is
\begin{equation}
\label{eq:deltat_spot}
\Delta t_{\rm spot} = \frac{\int \left[ 1 - \delta(t) + \delta_s(t) \right] (t-t_c)~dt}{\int \left[ 1 - \delta(t) + \delta_s(t) \right]~dt }
\approx \frac{\int \delta_s(t)~(t-t_c)~dt}{\int [1-\delta(t)] ~dt },
\end{equation}
where $t_c$ is the centroid of the idealized light curve. The simplification of the numerator is due to definition of $t_c$, and the simplification of the denominator assumes the perturbation is small. The spot anomaly $\delta_s(t)$ can be modeled as a triangular function of amplitude $A_s$, duration $T_s$ and midpoint $t_s$. For a spot smaller than the planet, the duration $T_s$ is approximately $(R_p/R_\star)T$, where $T$ is the time between the ingress and egress midpoints.  In such cases $T_s \ll T$, and Eqn.~(\ref{eq:deltat_spot}) simplifies to
\begin{equation}
\Delta t_{\rm spot} \approx
\frac{ \frac{1}{2} A_s T_s (t_s - t_c)}{(R_p/R_\star)^2 T},
\end{equation}
and for a spot anomaly at ingress or egress ($t_s-t_c \approx \pm T/2$),
\begin{equation}
\label{eq:timing-noise}
\Delta t_{\rm spot} \approx \pm \frac{A_{\rm s} T_{\rm s}}{4(R_p/R_\star)^2} \approx
(\pm 23~{\rm sec}) \left( \frac{A_s}{1500~{\rm ppm}} \right) \left( \frac{T_s}{0.4~{\rm hr}} \right),
\end{equation}
where the numerical factors are based on the observed WASP-4 parameters (see the next two sections and Table~\ref{tbl:spots}, giving the results of photometric spot modeling). The spot anomalies we identified have $A_s\approx 1500$~ppm, but if the very same spot had been crossed on the limb of the star rather than near the center of the disk, the anomaly would have been reduced by a factor of a 3-5 due to limb darkening and geometrical foreshortening, giving $A_s \approx$~300-500~ppm.  Such a small anomaly would not have been readily detected as a clear ``bump'' in our data, and according to Eqn.~(\ref{eq:timing-noise}) it would have produced timing noise of order 5-10~s, which is consistent with the excess scatter observed in the calculated transit midpoints.\footnote{We also used the photometric spot model described in \S~4 to confirm that the same spots that produced detectable anomalies could also produce timing noise of 5-10~s.  Specifically, we computed an idealized transit model $\delta(t)$ and added a spot model $\delta_s(t)$ based on the same spot parameters that were inferred from the actual data, but centered on the ingress rather than near midtransit. We then added Gaussian noise to mimic the actual data and fitted the resulting time series to derive the midtransit time. The offset was 8~s.}

We conclude that timing offsets due to starspot anomalies are a plausible explanation for some (and perhaps all) of the excess timing noise that was observed. Confirming the alternate hypothesis of gravitational perturbations would require the detection of a clear pattern in the residuals rather than just excess scatter (see, e.g., Holman et al.~2010), and is not possible with this relatively small number of data points. Table~\ref{tbl:params} gives the results for the reference epoch and orbital period, based on the 15-point fit, and with uncertainties based on the internal errors of the linear fit multiplied by $\sqrt{\chi^2/N_{\rm dof}}$, where $N_{\rm dof}$ is the number of degrees of freedom.

\begin{deluxetable}{lccc}
\tablecaption{Photometry of WASP-4 (Excerpt)\label{tbl:data}}
\tablewidth{0pt}

\tablehead{
\colhead{BJD$_{\rm TDB}$} & 
\colhead{Relative flux} &
\colhead{Uncertainty} &
\colhead{Airmass}
}

\startdata
  2454697.710091  &  1.00020  &  0.00067  &   1.083 \\
  2454697.710564  &  1.00047  &  0.00067  &   1.082 \\
  2454697.711039  &  0.99977  &  0.00067  &   1.081
\enddata 

\tablecomments{The time-stamp represents the Barycentric Julian Date at midexposure,
calculated based on the Julian Date with the code of Eastman et al.~(2010).
We intend for the rest of this table to be available online. \vskip 0.2in}

\end{deluxetable}

\begin{deluxetable*}{lccc}
\tabletypesize{\scriptsize}
\tablecaption{Midtransit times and apparent transit depths of WASP-4b\label{tbl:individual}}
\tablewidth{0pt}

\tablehead{
\colhead{Date} & \colhead{Epoch} & \colhead{Midtransit time (BJD$_{\rm TDB}$)} & \colhead{Transit depth $(R_p/R_\star)^2$}
}

\startdata
2008~Aug~19 & $0$      & $2454697.798151 \pm 0.000056$ &  $0.02436\pm 0.00017$ \\
2008~Oct~09 & $38$     & $2454748.651175 \pm 0.000049$ &  $0.02370\pm 0.00015$ \\
2009~Aug~02 & $260$    & $2455045.738643 \pm 0.000054$ &  $0.02402\pm 0.00013$ \\
2009~Aug~06 & $263$    & $2455049.753274 \pm 0.000066$ &  $0.02353\pm 0.00014$ \\
2009~Aug~10 & $266$    & $2455053.767816 \pm 0.000053$ &  $0.02373\pm 0.00014$ \\
2009~Sep~26 & $301$    & $2455100.605928 \pm 0.000061$ &  $0.02379\pm 0.00014$
\enddata

\end{deluxetable*}

\begin{deluxetable*}{lcc}
\tabletypesize{\scriptsize}
\tablecaption{System Parameters of WASP-4\lowercase{b}\label{tbl:params}}
\tablewidth{0pt}

\tablehead{
\colhead{Parameter} & \colhead{Value} & \colhead{68.3\% Conf.~Limits} 
}

\startdata
Reference epoch~[BJD$_{\rm TDB}$]                      & $2454697.798226$  &  $\pm 0.000048$      \\
Orbital period~[days]                                 & $1.33823187$      &  $\pm 0.00000025$    \\
Planet-to-star radius ratio, $R_p/R_\star$\tablenotemark{a}             & $0.1544$         &  $\pm 0.0009$  \\
Orbital inclination, $i$~[deg]                        & $88.80$           &  $-0.43$, $+0.61$    \\
Scaled semimajor axis, $a/R_\star$                     & $5.482$           &  $-0.022$, $+0.015$    \\
Transit impact parameter, $b = a\cos i/R_\star$        & $0.115$           &  $-0.058$, $+0.040$   \\
Transit duration~[hr]                                 & $2.1585$          &  $-0.0036$, $+0.0038$   \\
Transit ingress or egress duration~[hr]               & $0.2949$          &  $-0.0025$, $+0.0030$   \\
Linear limb-darkening coefficient, $u_1$                & $0.305$           &  $\pm 0.023$   \\
Quadratic limb-darkening coefficient, $u_2$             & $0.173$           &  $\pm 0.089$   \\
Mass of the star, M$_\star$~[M$_{\odot}$]\tablenotemark{b}                  & $0.92$    & $\pm 0.06$ \\
Semimajor axis~[AU]                                   & $0.02312$         &  $\pm 0.00033$  \\
Radius of the star, $R_\star$~[R$_{\odot}$]        &  $0.907$  & $-0.013$, $+0.014$  \\
Radius of the planet, $R_p$~[R$_{\rm Jup}$]           &  $1.363$         &  $\pm 0.020$
\enddata

\tablecomments{The quoted result for each parameter represents the median of the {\it a posteriori} probability distribution derived from the MCMC method and marginalized over all other parameters. The confidence limits enclose 68.3\% of the probability, and are based on the 15.85\% and 84.15\% levels of the cumulative probability distribution.}

\tablenotetext{a}{Represents the weighted average of the 6 different results for the planet-to-star radius ratio. The quoted uncertainty in the final value is
the standard deviation of these 6 results.}

\tablenotetext{b}{The stellar mass of $0.92\pm 0.06$~$M_\odot$ was adopted based on the analysis of Winn et al.~(2009), and used to derive the following three parameters. \vspace{0.1in}}

\end{deluxetable*}

\section{Spot model: photometric}
\label{sec:model-photometric}

A central question for our study is whether each pair of starspot anomalies was caused by occultation of the {\it same} spot. One issue is whether a spot could last long enough to be occulted twice. The two anomalies seen in our data were separated in time by 47~days, and the two anomalies in the Southworth et al.~(2009) data were separated by 31 days. On the Sun, individual spots last from hours to months, with a lifetime proportional to size following the so-called GW rule (Gnevyshev 1938, Waldmeier 1955): $A_0 = WT$, where $A_0$ is the maximum spot size in MSH (micro-solar hemispheres), $T$ is the lifetime in days, and $W = 10.89\pm 0.18$ (Petrovay \& Van Driel-Gesztelyi 1997). The amplitudes of the WASP-4 anomalies are $\approx$1500~ppm, suggesting that the spot area is of order 2000~MSH and giving a GW lifetime of 180 days. However, the application of this rule to WASP-4 requires an extrapolation, since the implied spot size is several times larger than most sunspots (Solanki 2003). Henwood et al.~(2010) studied larger spots, and found them to follow the same rule, but with a relatively small sample size.

From this perspective it is plausible that each pair of anomalies represents two passages of the planet over the same spot. However, the spot that was observed with Magellan is not likely to be the same spot that was observed by Southworth et al.~(2009) because those two groups of observations were conducted one year apart. This conclusion is borne out by the modeling described below.

Another issue is whether the {\it amplitudes} and {\it durations} of both events in a pair are consistent with passage over a single spot. A photometric spot model will make specific predictions regarding the observable anomalies, based on the stellar limb-darkening law, the geometrical foreshortening of the spots and the orbital velocity of the planet. We are reluctant to take such a model too seriously, given the unknown shape of the spot and the potential for time variations in its shape and intensity. In the case of the Sun, spots reach their maximum size within a few days and then shrink with time at a rate of about 30 MSH~day$^{-1}$ (Solanki 2003). Another complication is that spots can migrate to different latitudes, although for the Sun this migration amounts to fewer than 5 degrees (Henwood et al.~2010). Nevertheless we used a model with static spot properties to perform a consistency check on the hypothesis that the same spot was occulted twice.

The orientation of the star was parameterized by $\lambda$, the sky-projected spin-orbit angle, and $i_s$, the inclination of the stellar rotation axis with respect to the line of sight, using the coordinate system of Ohta et al.~(2005). The visible hemisphere of the star was pixellated with a $241 \times 241$ Cartesian grid (enough to allow for fast computations with tolerable discretization error), and the pixels were assigned intensities using a quadratic limb-darkening law.  The planet's trajectory was computed from the known orbital parameters, and zero intensity was assigned to those pixels covered by the planet's silhouette.  The spot was taken to be a circle of lower intensity on the stellar photosphere, and its geometrical foreshortening was taken into account in assigning intensities to the affected pixels. The intensity distribution within the spot was taken to be a Gaussian function with a truncation radius equal to three times the standard deviation of the distribution. (We also tried modeling spots with a constant intensity, which gave qualitatively similar results.) The model had seven adjustable parameters: the stellar orientation angles $\lambda$ and $i_s$, the rotation period of the spot, the spot intensity and radius, and the initial longitude and latitude of the spot at the time of the first anomaly.

\begin{figure*}[ht]
\begin{center}
\leavevmode
\hbox{
\epsfxsize=6.0in
\epsffile{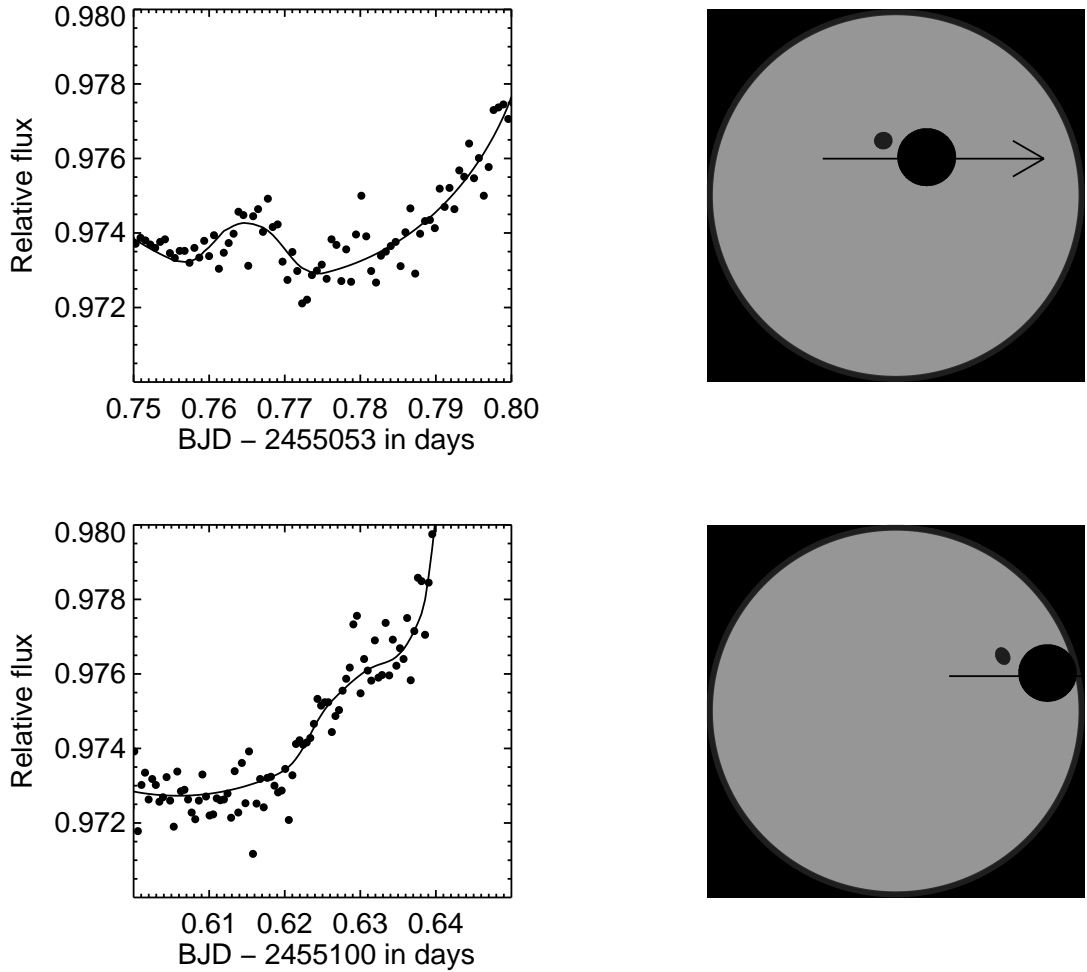}}
\end{center}
\vspace{-0.20in}
\caption{A closer look at the spot anomalies. {\it Left}: The relevant portion
of the light curves, along with the best fitting transit~$+$~spot model.
{\it Right}: Corresponding positions of the spot on the transit chord.}
\vspace{0.0in}
\label{fig:spots}
\end{figure*}

For simplicity we studied the well-aligned case $\lambda=0^\circ$, $i_s = 90^\circ$. The best-fitting model is displayed in Figure~\ref{fig:spots}. The amplitudes and durations of the anomalies are fitted well, and the optimized rotation period is 22.2~days, i.e., the second anomaly was observed slightly more than two complete rotations after the first anomaly. This is within the broad range of periods, 20-40~days, that is expected for a main-sequence G7 star (see, e.g., Barnes 2007, Schlaufman 2010). In addition, this value for the rotation period agrees with the value that can be estimated from the sky-projected rotation rate $v\sin i_s$ and the stellar radius $R_\star$ according to
\begin{equation}
P_{\rm rot} \approx \frac{2\pi R_\star}{v\sin i_s}~\sin i_s = (21.5 \pm 4.3~{\rm days})~\sin i_s,
\end{equation}
where we have used $v\sin i_s = 2.14\pm 0.37$~km~s$^{-1}$ from the work of Triaud et al.~(2010), and $R_\star = 0.907\pm 0.014$~$R_\odot$ from our analysis.

In the best-fitting model, the spot's intensity profile has a maximum contrast of 32\% with respect to the surrounding photosphere.  Modeling both the photosphere and the spot as blackbodies, and using $T_{\rm eff} = 5500$~K for the photosphere (Wilson et al.~2010), the corresponding spot temperature is 4900~K.  The spot radius is $0.05~R_\star$, implying that it is significantly smaller than the planet ($0.15~R_\star$). The spot radius and intensity contrast are highly correlated; only their product is well determined.

The fit seems reasonable in all respects and correctly predicts the nondetection of anomalies during the first and second nights of observations. Other local minima in $\chi^2$ can be found involving a larger number of rotations between anomalies, with $P_{\rm rot} =15.1$ or $11.4$~days, but these give $\Delta\chi^2 \approx 10$ relative to the global minimum and rotation periods outside of the expected range.  A similar analysis of the Southworth et al.~(2009) data shows that the spot is about the same size, and gives possible rotation periods of 25.5~days and 14.0~days, of which the former is closer to the Magellan result and to the expected value.

We concluded from this exercise that each dataset (ours and that of Southworth et al.~2009) is consistent with a single spot and a star that is well-aligned with the orbit. We decided not to pursue the implications of this photometric starspot model further, given that the simplifying assumptions (such as a circular, unchanging spot) lead to more significant uncertainties than the photometric uncertainties. In particular, the results for $\lambda$ and its uncertainty would depend on the assumed shape of the spot, because the planet trajectories with $\lambda \neq 0$ could graze the spot at different angles during each encounter. Instead we used a simplified model constrained almost exclusively by the timings of the anomalies, as described in the next section.

\vspace{0.2in}
\section{Spot model: geometric}
\label{sec:model-geometric}

The recurrence of the anomaly at a later phase of the transit favors the configuration where the orbital angular momentum and the axis of rotation of the star are aligned, because in such a situation the trajectories of the spot on the surface and the planet would be almost parallel. The purpose of the geometric model described in this section is to quantify this statement, based only the observed times of the anomalies, without attempting to model complicated and largely irrelevant aspects of the situation such as the full range of possibilities for the spot size, intensity, and possible nonuniform motions.

To measure the times and gain an appreciation of the statistical significance of each feature, we used a simple triangular model for each anomaly.  The triangular model is overplotted upon the residuals in Figure~\ref{fig:magtrans}.  Table~\ref{tbl:spots} gives the results for the parameter values.  As shown in the last few rows of that table, the first three spot anomalies (the two Magellan anomalies, and the first Southworth et al.\ anomaly) are detected with relatively high confidence.  The spot model includes 3 extra free parameters, and improves the fit by $\Delta\chi^2 = 85$, 34 and 25, for each of the first three transits, as compared to the best-fitting model with no spots. The fourth is marginal, with $\Delta\chi^2 = 8$.\footnote{All of these comparisons took time-correlated noise into account, in the sense that $\chi^2$ was computed assuming flux uncertainties that have been enlarged by the red-noise factor $\beta$.  The number of data points and number of degrees of freedom for each case are given in Table~\ref{tbl:spots}.} The weaker amplitude of the fourth event is consistent with the spot model, as the anomaly occurred near the egress where limb darkening and geometrical foreshortening both reduce the amplitude of the photometric effect. However, it remains possible that the ``anomaly'' is a spurious statistical detection.

Next we defined a likelihood function for $\lambda$ and $i_s$, given the observed times of anomalies as well as the observed time ranges of nondetections. The basic idea is to assume that the spot is located within the planet's shadow at the time of the first anomaly, and then compute the position of the spot at the other relevant times for a given choice of the parameters $\{\lambda, i_s, P_{\rm rot}\}$ (a purely geometric calculation). The model is rewarded for producing spot-planet coincidences at the appropriate times, and penalized for producing coincidences at inappropriate times. Each of the two spots---the one observed in 2008, and the one observed in 2009---is given an independent value of $P_{\rm rot}$ to allow for possible differential rotation or peculiar motions of the spots (see Section \ref{sec:discussion} for discussion). A further constraint is imposed to enforce agreement with the spectroscopic determination of $v\sin i_\star$ by Triaud et al.~(2010). Mathematically, we used a likelihood $\exp(-\chi^2/2)$ with
\begin{eqnarray}
  \chi^2 ( P_{{\rm rot}, 1}&,& P_{{\rm rot}, 2}, \lambda, i_s) = \nonumber \\
  \sum_{j=1}^2 \left(\frac{d_j}{R_p/2}\right)^2 & + & \left[ \frac{(2 \pi R_s/P_{{\rm rot},j}) \sin{i_s} - 2.14 }{0.37} \right]^2  + {\rm NDP},
 \end{eqnarray}
where $j$ is the index specifying one of the two anomalies, and $d$ is the distance on the stellar disk between the center of the planet and the center of the spot. Thus, high likelihoods are assigned to spot-planet coincidences within 0.5~$R_p$ at the correct times. This factor is based on the estimation of the size of the spot given by the photometric model, and it would require modification if the spot were bigger than the
planet. The factor NDP is the nondetection penalty: models that produce spot-planet coincidences at times when they were not observed are ruled out by incrementing $\chi^2$ by 1000 (an arbitrary number chosen to be large enough to exact a severe penalty). Based on our studies of the amplitude of the spots with the more sophisticated model of Section 3, the nondetection penalty was only applied for coincidences within 0.9~$R_\star$ of the center of the stellar disk. For the outer 0.1~$R_\star$ (near the limb) the combined effects of limb-darkening and foreshortening would have made such an anomaly undetectable.

We used an MCMC algorithm, with the Gibbs sampler and Metropolis-Hastings criterion, to sample from the posterior probability distribution for the parameters, with uniform priors on $\lambda$ and $\cos i_s$ (i.e., isotropic in the stellar orientation).  We restricted $|\lambda| < 90^\circ$, given the finding of Triaud et al.~(2010) that the orbit is prograde, based on the Rossiter-McLaughlin effect. Given our finding of multiple minima in the photometric model (Section~\ref{sec:model-photometric}), we also performed a dense grid search in the two-dimensional space of $P_{{\rm rot},1}$ and $P_{{\rm rot},2}$. This identified four relevant local minima with periods $>$10~days (smaller periods were rejected as unlikely for a star of the observed mass and age). A Markov chain was initiated from each of these 4 minima.

\begin{figure*}[ht]
\begin{center}
\leavevmode
\hbox{
\epsfxsize=6.0in
\epsffile{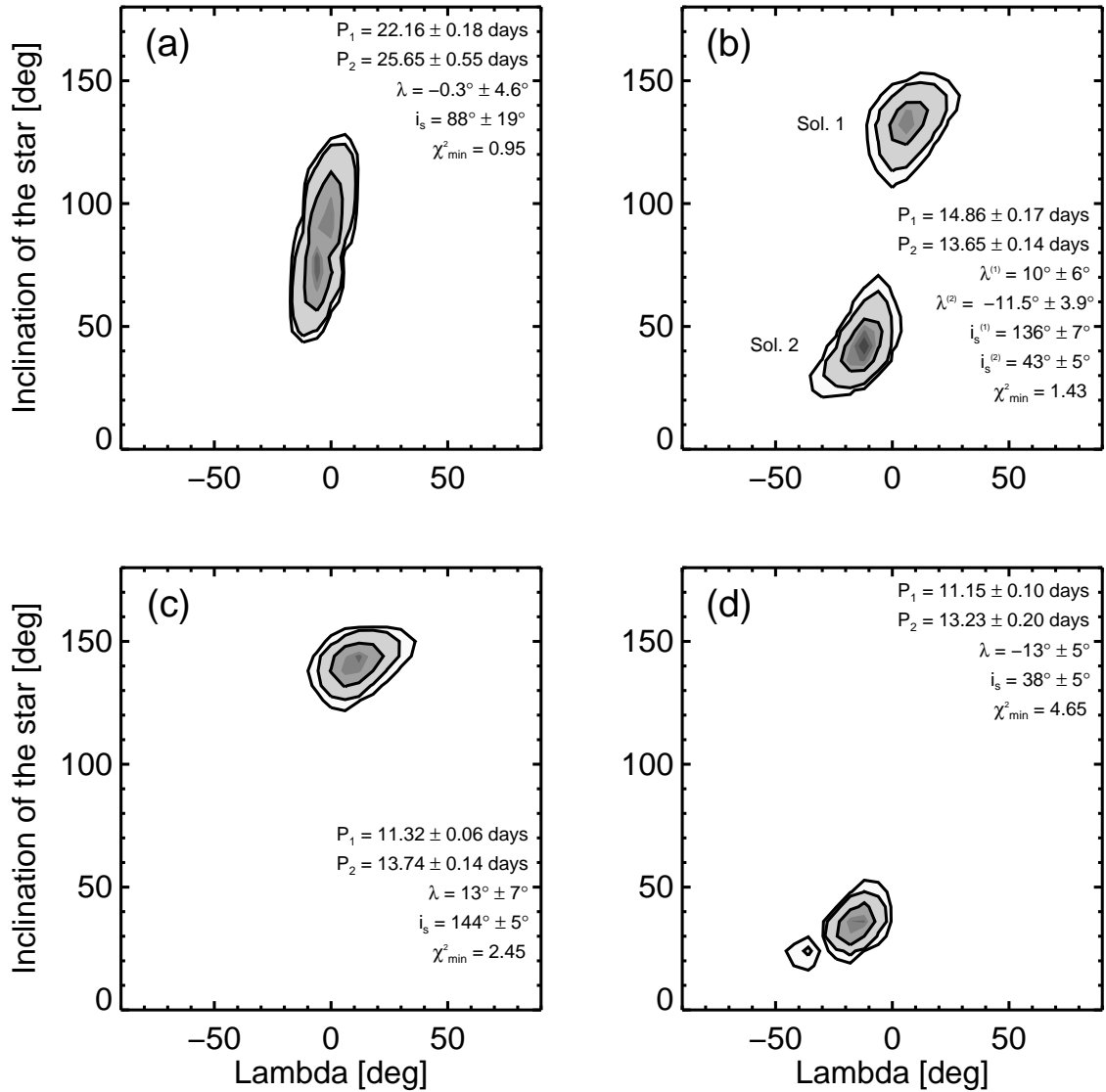}}
\end{center}
\vspace{0.00in}
\caption{Combined probability distribution of $\lambda$ and $i_s$ four all four different solutions. Printed on each plot are
the parameter values and uncertainties.}
\vspace{0.0in}
\label{fig:chi2plots}
\end{figure*}

\begin{figure*}[ht]
\begin{center}
\leavevmode
\hbox{
\epsfxsize=6.0in
\epsffile{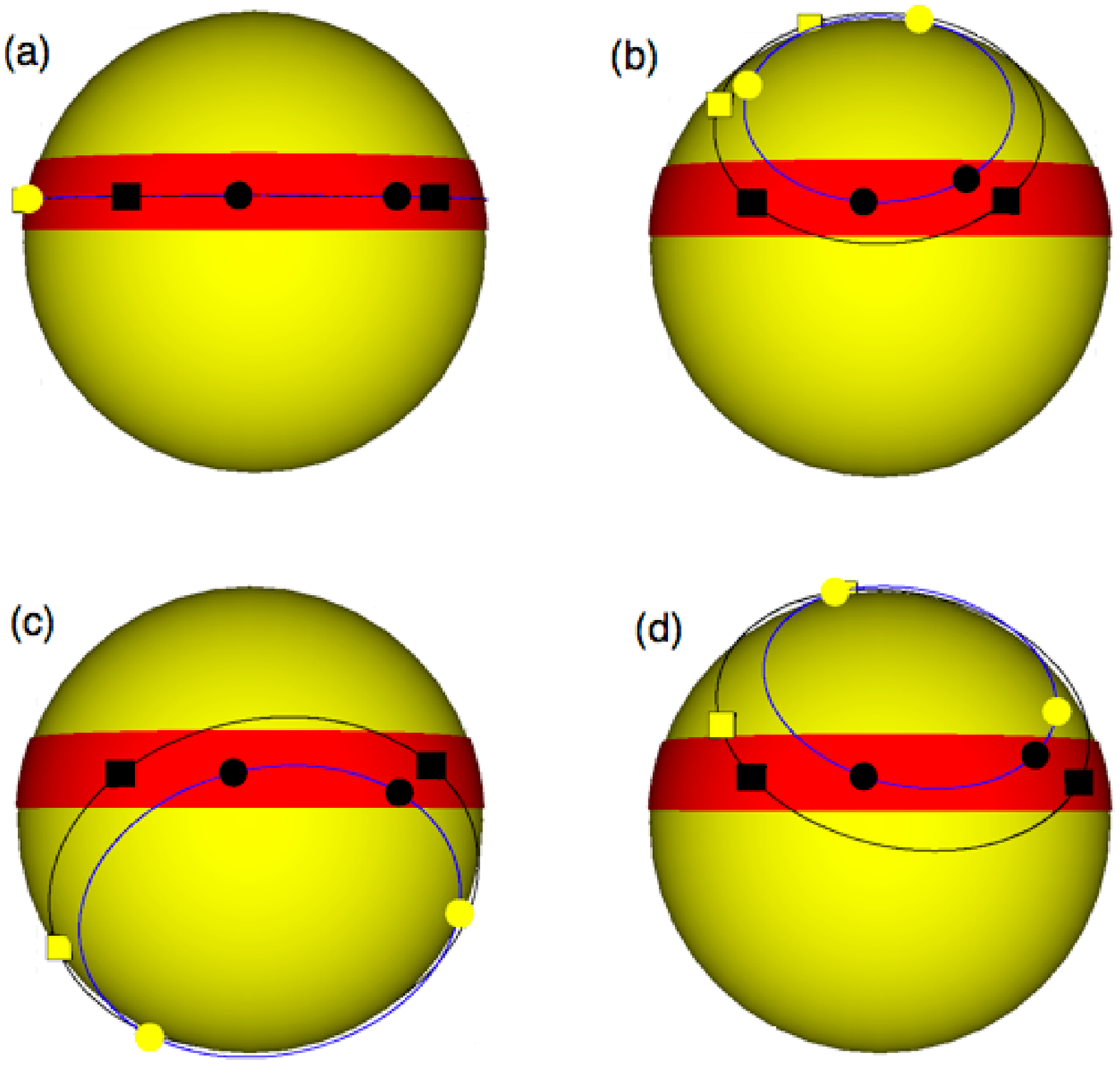}}
\end{center}
\vspace{0.00in}
\caption{Visualizations of the four different solutions. Circles represent the position of the spot during the transits
we observed, and squares represent the same for Southworth's observations. The dark symbols represent detections and the light symbols represent nondetections. The shaded area of the star represents the transit chord. In the case of the upper right panel, we have plotted the geometry corresponding to only one of the two possible values of $i_s$ shown in Figure~\ref{fig:chi2plots}b (specifically the smaller value).}
\vspace{0.0in}
\label{fig:3dsphere}
\end{figure*}

Figure~\ref{fig:chi2plots} shows the 2-d probability distribution for $\lambda$ and $i_s$
for all four possible solutions, after marginalizing over the rotation periods. The first thing to notice is that small values of $\lambda$ are favored in all cases, while $i_s$ is poorly constrained.  The completely aligned case (upper left corner of the panel) is the global minimum, with $\chi^2=0.95$, but none of the other solutions can be firmly ruled out.

These results can be understood by visualizing the various solutions, as we have done in Figure~\ref{fig:3dsphere}.  The four different configurations shown in that figure correspond to the four local minima.  (One of the minima actually gave a bimodal distribution, as shown in the upper right panel of Figure~\ref{fig:chi2plots}; for that case Figure~\ref{fig:3dsphere} shows the small-$i_s$ solution.)  The upper right panel shows the completely aligned case.  This type of solution is always possible whenever two anomalies from the same spot are observed at different transit phases, unless it is ruled out by the nondetection of anomalies that should be present in other light curves.  In our case, the model predicts an anomaly during the ingress of the $E=0$ transit, and also right at the ingress of the $E=263$ transit. Neither of these anomalies would have been detectable in our data. The other three panels show how an appropriate combination of $\lambda$ and $i_s$ causes the trajectory of the spot to move outside the transit chord and then back inside in time for the second anomaly.

The well-aligned case is favored not only because of the lower $\chi^2$, but also because the corresponding rotation periods (22 and 26~days) are within the expected range of 20--40 days, as opposed to the shorter periods associated with the other solutions.  One could also argue that for any observational campaign involving only a few transits, the detection of multiple spot anomalies is {\it a priori} more likely for a well-aligned system than for a misaligned system, because in the former case the spot spends a much larger fraction of the time on the transit chord. For simplicity, though, we report a determination of $\lambda$ based on the simple concatenation of all the Markov chains corresponding to the four local minima, giving $\lambda = -1^{+14}_{-12}$~degrees.

\begin{deluxetable*}{lcccc}

\tabletypesize{\scriptsize}
\tablecaption{Characterization of the spots \label{tbl:spots}}
\tablewidth{0pt}

\tablehead{
\colhead{} &
\colhead{2009 Aug 10 } &
\colhead{2009 Sep 26}&
\colhead{2008 Aug 23}&
\colhead{2008 Sep 23 }
}

\startdata
Amplitude (ppm) & 1790 & 1470 & 2400 & 1190 \\ 
Duration (hours) & 0.34 & 0.38 & 0.56 & 0.54 \\ 
Time of event (BJD$_{\rm TDB}$) & 2455053.7658 & 2455100.6288 & 2454701.7938 & 2454732.6172 \\ 
Epoch $E$ & 266 & 301 & 3 & 26 \\ 
RMS residual (ppm) & 523 & 580 & 765 & 722 \\ 
Number of data points                    & 365 & 355 & 126 & 88  \\ 
$\chi^2$ ($N_{\rm dof}$) for no-spot model & 435 (358) & 220 (348) & 102 (119) &  69.4 (81) \\ 
$\chi^2$ ($N_{\rm dof}$) with spot model   & 350 (355) & 186 (345) & 77 (116) &  61.4 (78) \\ 
$\Delta\chi^2$ & 85 & 34 & 25 & 8
\enddata

\tablecomments{Parameters of the best fitting models to the residuals of the four different spot events. Note that $\chi^2$ was computed after enlarging the flux uncertainties by the red-noise factor $\beta$ described in \S~\ref{sec:starspots_and_system_parameters}.}

\end{deluxetable*}

\section{Summary and discussion}
\label{sec:discussion}

In this paper, we report the observations of four new transits of the WASP-4b planet, observations that lead to a significant improvement on the errors of the system parameters and the transit ephemerides. Short-lived photometric anomalies, transit timing variations and transit depth variations were all observed, all of which can potentially be explained by the effects of starspots. In particular we have interpreted the photometric anomalies as occultations of starspots by the planet. We have described a simple method for assessing the orientation of a star relative to the orbit of its transiting planet through the analysis of spot occultations. This method has certain advantages and disadvantages compared to observations of the RM effect, the main method for such determinations.

On the positive side, the spot method works well for slowly-rotating stars, for which the RM amplitude is smallest. The spot method also has no particular problem with low impact parameters, unlike the RM effect. These two factors help to explain why the spot method gives tighter constraints on $\lambda$ than did the RM observations of Triaud et al.~(2010), for the case of WASP-4.  The spot method requires that the star be moderately active. This too is complementary to RM observations, which rely on precise Doppler spectroscopy and are hindered by stellar activity. In addition, the spot method is photometric rather than spectroscopic, and as such it does not require a high-resolution spectrograph nor special efforts to achieve accurate radial-velocity precision.

On the negative side, many transits must be observed to have a reasonable chance of detecting multiple anomalies, and to be sure that multiple anomalies are caused by a single spot, rather than distinct spots. In the case of WASP-4, a few more transit observations during the summers of either 2008 or 2009 could have allowed for a more secure validation of the single-spot hypothesis, and removed the four-way degeneracy of the resulting constraints on the stellar orientation.  Furthermore, spots are not well-behaved deterministic entities: they have irregular shapes that form and dissolve, governed by poorly understood physical principles.

Regarding that subject, it is interesting to note that all four of the solutions shown in Figure~\ref{fig:chi2plots} involve slightly but significantly different rotation periods for the spot seen in 2008 as compared to the one seen in 2009.  This could be a sign of differential rotation.  Assuming WASP-4 has $\lambda=0^\circ$ and has the same differential rotation profile as the Sun, spots on the top and bottom of the transit chord would have periods differ by 10\%, as compared to the 10-15\% differences seen in our model results.  Thus, differential rotation is a realistic possibility.  Another contributing factor may be peculiar motions of spots, i.e., motions of the spot relative to the surrounding photospheres. On the Sun, individual spots at a given latitude are observed to have rotation periods differing by a few percent (Ru{\v z}djak et al.~2005).

For WASP-4, the small value of $\lambda$ is further evidence that this is a low-obliquity system.  Such findings have been interpreted as constraints on the process of planet migration: the mechanism that brought this gas giant planet from its birthplace (presumably a few AU) to its close-in orbit. Low obliquities are suggestive of disk migration, in which the orbit shrinks due to tidal interactions with the protoplanetary gas disk; while large obliquities would favor theories in which close-in orbits results from gravitational interactions with other bodies followed by tidal dissipation. The complicating factor of tidal reorientation was thought to be negligible, but this possibility was recently raised by Winn et al.~(2010a) as a possible explanation for the tendency for high-obliquity stars to be ``hot'' and low-obliquity stars to be ``cool'', with a boundary at around 6250~K.  Here we will not remark further on the theory underlying this hypothesis, but simply note that WASP-4 conforms to the empirical pattern, as a cool and low-obliquity system.

Looking forward, an opportunity exists to implement this method for other systems using the data from the {\it CoRoT} and {\it Kepler} space missions. The CoRoT-2 system in particular has a highly spotted star (see, e.g., Silva-Valio et al.\ 2010, Silva-Valio \& Lanza 2011) for which our method might be applicable, although the spots are so numerous and influential on the light curve that more complex models may be necessary. {\it Kepler} employs a 1m space telescope to monitor 150,000 stars with photon-limited precision down to level of $\approx$10 parts per million (Borucki et al.~2010, 2011). The data released in February 2011 displays a limiting precision of about 10~ppm in 6~hr combined integrations at {\it Kepler} magnitude 10 (approximately $r=10$), and a limiting precision of about 100~ppm for a more typical target star magnitude of 15. Besides high precision, the great advantage of the space missions is nearly-continuous data collection. For a system resembling WASP-4, {\it Kepler} would observe hundreds of consecutive transits, resulting in much greater power to track individual spots.  Furthermore, the brightness variations observed outside of transits will allow for an independent estimate of the stellar rotation period, as well as additional constraints on spot longitudes. A potentially serious problem with the application to {\it Kepler} is that most stars are observed with a cadence of 30~min, which may be too long to pin down the times of starspot anomalies with the required precision. A subset of targets are observed at the much more favorable cadence of 1~min. Already there is one transit-hosting star in the {\it Kepler} field, HAT-P-11, that is being observed with 1~min cadence and will assuredly yield interesting results, as $\lambda$ was found to be approximately $100^\circ$ by Winn et al.~(2010b) and Hirano et al.~(2011), and the star has long-lived, sizable spots (Bakos et al.~2010).

\acknowledgements We thank the anonymous referee for a comprehensive and helpful report on the manuscript.  We gratefully acknowledge support from the NASA Origins program through awards NNX09AD36G and NNX09AB33G, and the MIT Class of 1942. R.S.\ received financial support through the ``la Caixa'' Fellowship Grant for Post-Graduate Studies, Caixa d'Estalvis i Pensions de Barcelona ``la Caixa'', Barcelona, Spain. J.A.C.\ acknowledges support for this work by NASA through Hubble Fellowship grant HF-51267.01-A awarded by the Space Telescope Science Institute, which is operated by the Association of Universities for Research in Astronomy, Inc., for NASA under contract NAS 5-26555.

\end{document}